# Boundary layer instability control in the unsteady cloud cavitating flow

Ebrahim Kadivar and Ould el Moctar

Institute of Ship Technology, Ocean Engineering and Transport Systems, University of Duisburg-Essen, Duisburg, Germany
E-mail: ebrahim.kadivar@uni-due.de

**Abstract**
In this article, we propose a passive boundary layer control method to control the vortex structure of the cavity on the suction side and wake region of the CAV2003 benchmark hydrofoil. This method may be used in different applications such as marine, turbomachinery and hydraulic machinery. First, we used a hybrid URANS model for turbulence to simulate the 3D unsteady cloud cavitating flow and validated it based on experimental data. We performed the numerical simulations using open source code OpenFOAM and an Euler-Euler cavitation model. Second, we studied the effect of passive boundary layer control method on vortex structure on the suction side of the hydrofoil and in wake region. We showed that this control method may influence the boundary layer structure on the hydrofoil surface and also near the trailing edge. Using this technique the pressure distribution and the fluctuating part of the velocity field on the hydrofoil surface were modified over the chord length. This method induced a stabilization of the boundary layer and delay its separation. Therefore a significant reduction in cavitation-induced vibration may be expected.

## 1. Introduction

In the past decade, many researchers performed the simulation of cavitation phenomenon over submerged bodies such as hydrofoils, propellers, pump and turbine blades by experimental works and numerical methods. Recently researchers have tried to develop passive methods to control cavitation and reduce undesirable behaviors of cavitation. Most of the control methods which were investigated by the researchers focused on the control of the steady partial cavitation and super-cavitation, which are a quasi-stable type of cavitation. The effects of the passive control at the cavitation structure was investigated using different cavitators, [1-5]. Their results illustrated that the shape and the wedge angle of the cavitators affects significantly the structure and type of supercavitation. Various studies on venturi-type geometries shown that passive control method based on distributed roughness influence sheet cavitation. Delgosha et al. [6] studied the effect of the surface roughness on the dynamics of sheet cavitation on a two-dimensional foil section. They reported that the roughness in the downstream end of the sheet cavity may change the arrangement of the cavitation cycle. Zhang et al. [7] mentioned that the roughness of the surface may control the development of the cavity. The effect of leading edge roughness on the boundary layer and transition was studied by Kerho et al. [8] and Dryden [9]. They showed that the roughness triggered the transition at the leading edge. The aspects of cavitation instability were investigated by Kawanami et al [10], Wade et al. [11], Arndt et al. [12] and Watanabe [13]. They showed that the attached cavity developing on the suction side of a hydrofoil may be unstable and causes a significant vibration when the length of the attached cavity exceeds about 70% of the chord length. They compared different types of unsteady cavitation and showed that the large cloud cavitation oscillates between partial cavity and supercavity with a relatively low Strouhal number in the range of 0.15 to 0.3 when the maximum cavity length is greater than 75% of the chord length. Passive control method in unsteady cloud cavitation regime were studied by different authors. Danlos et. al [14] investigated the effects of the surface condition of a venturi profile experimentally. They showed that the roughness surfaces can suppress the shedding of the unsteady partial cavitation. Ausoni et al. [16] investigated the hydrofoil roughness effects on von Karman vortex shedding in cavitation free regime. Other studies deal with the boundary layer and cavitation. Arakeri [17] and Katz [18] shown that after the laminar separation phenomenon the sheet cavity can be stabilized by the separated bubble. The relation between cavitation and laminar separation of the boundary layer was investigated by Franc and Michel [19]. They showed that the cavity does not detach from the body at the minimum pressure point. The relation between the transition of boundary layer and generation of the unsteady cloud cavitation was studied by Avellan et al. [20]. The literature review reveals that most of the previous investigations were focused on the effects of the passive control method on the sheet cavitation dynamics and vortex shedding in non-cavitating flow. Most of the passive control techniques were performed on the venturi-type geometries. Methods to control unsteady cloud cavitation which may damage the solid walls of the immersed bodies and causes different destructive effects such as noise and vibration have been developed and applied by few authors. However, details related to boundary layer instability control affect the unsteady cloud cavitation and the vortex structure of the cavity on the surface of immersed bodies were not studied yet. The goal of this work is to present a passive method to control the local boundary layer instabilities on the suction side of CAV2003 benchmark hydrofoil.

## 2. Methodology

We adapted this idea of the passive control from vortex generators (VGs) which are common in boundary layer control around airfoils in aerospace engineering applications. Because of their small size and high performance, the VGs are one of the effective methods to control flow separation on airfoils, Gad-el-Hak [21]. Using vortex generators the free-stream flow with high fluid momentum can be transferred into the vicinity of the wall surface of the hydrofoil with low energy fluid. The created vortices bring the fluid with higher kinematic energy to withstand a pressure rise before the separation phenomenon occurs. This method may be used in hydrodynamic applications to delay or suppress the boundary layer instabilities and flow separation on the suction side of hydrofoils under non-cavitating and cavitating conditions. In this work we used a wedge-type called cavitating-bubble generator (CG) located on the suction side of the hydrofoil where it is expected that the boundary layer becomes instabil. The view of the same shape of wedge-type CGs located in four different positions on the suction side of hydrofoil was shown in Fig. 1. Our investigations on the hight of the CGs show that it should be small enough so that it does not have a significant effect on the hydrodynamic performance of the hydrofoil. First, we estimated the location of inception point on the suction side of the hydrofoil without CGs. Second, we inserted a CG at different locations of the boundary layer in front and behind the inception point on the suction side of hydrofoil near the leading edge. This leads to find a proper location of CG with regard to the reduction of the highest amplitude corresponding to the cavitation shedding. For more details see [15], [24] and [25].

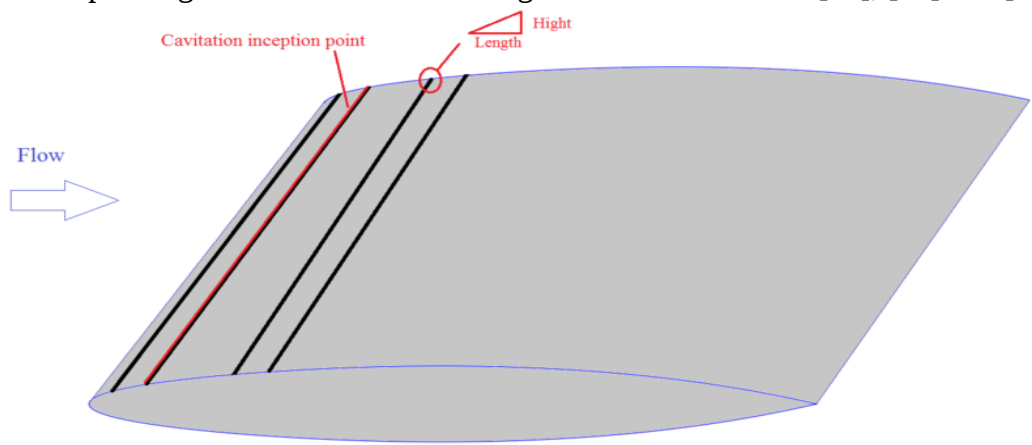

**Figure 1.** View of four wedge-type CGs located in four different positions on the suction side of hydrofoil, (black line). Position of cavitation inception point without CGs, (red line).

## 3. Numerical procedure

For the simulations of this study, we used the open source code OpenFOAM to solve cavitationmodel based on Euler-Euler solver. We applied the Schnerr-Sauer cavitation model. We used transport equation of volume fraction (TEM) to model the phases distribution based on the VOF method. We used a hybrid URANS model for turbulence model and solved the pressure-velocity coupling by a PIMPLE algorithm. The initial conditions and reference values for our simulation is shown in Table 1. We generated coarse, mid-size and fine grids. Effect of spatial discretization on time-averaged lift $C_l$, time-averaged drag coefficients $C_d$, maximum length of the attached cavity to the chord length $l/l_{ref}$ and Strouhal number based on the chord $St_c = f \times l_{ref}/V_{ref}$ are shown in Table 2 where $f$, $V_{ref}$, $l$ and $l_{ref}$ are the cavity self-oscillation frequency, reference velocity, maximum length of the attached part of the cavity and chord of the hydrofoil, respectively. The time step for the unsteady computation was set to 4e-6 s for a Courant number less than 1. For the mid-size grid we simulated the flow using two different time steps, 4e-6 s and 1e-5 s. The results showed that the the difference between these two time steps is not significant. The mesh near the wall of the test body is well refined to ensure the non-dimensional normal distance from the wall. The value of y + at the wall surface of hydrofoil was about 1.

**Table 1.** Initial conditions and reference values

| | |
|---|---|
| $\alpha = 7°$ | $P_v = 2000 Pa$ |
| $l_{ref} = 0.1m$ | $V_{ref} = 6.0 m/s$ |
| $T_{ref} = l_{ref}/V_{ref} = 0.0167s$ | $\rho_{ref} = 1000 kg/m^3$ |

**Table 2.** Comparison of the time-averaged lift and drag coefficients, maximum length of the attached cavity to the chord length and strouhal numbers at cavitation number $\sigma = 0.8$

| Mesh size | $\bar{C}_l$ | $\bar{C}_d$ | $l/l_{ref}$ | $St_c$ |
|---|---|---|---|---|
| coarse | 0.48 | 0.075 | 68% | 0.107 |
| mid-size | 0.44 | 0.073 | 70% | 0.11 |
| fine | 0.43 | 0.072 | 70% | 0.11 |
| Delgosha-Simulation [23] | 0.45 | 0.07 | 70% | 0.108 |
| Delgosha-Experiment [23] | - | - | 70% | 0.15 |

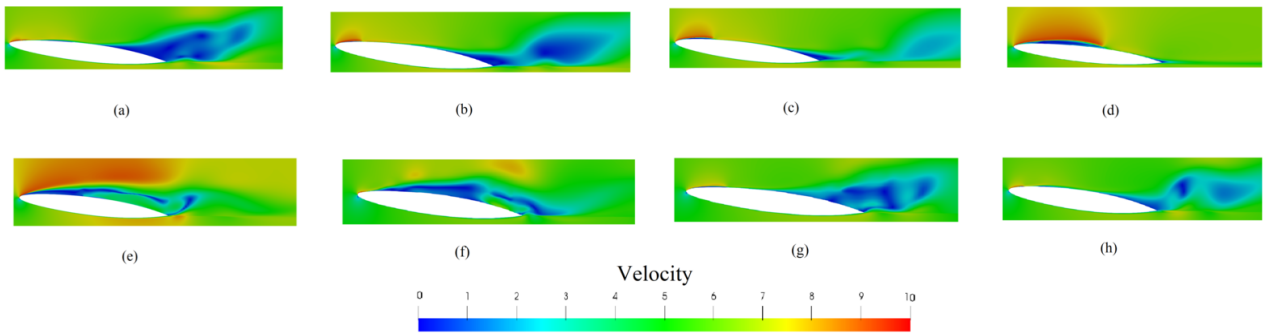

**Figure 2.** Instantaneous velocity structures around the hydrofoil without CGs. The time step between images is 19 ms.

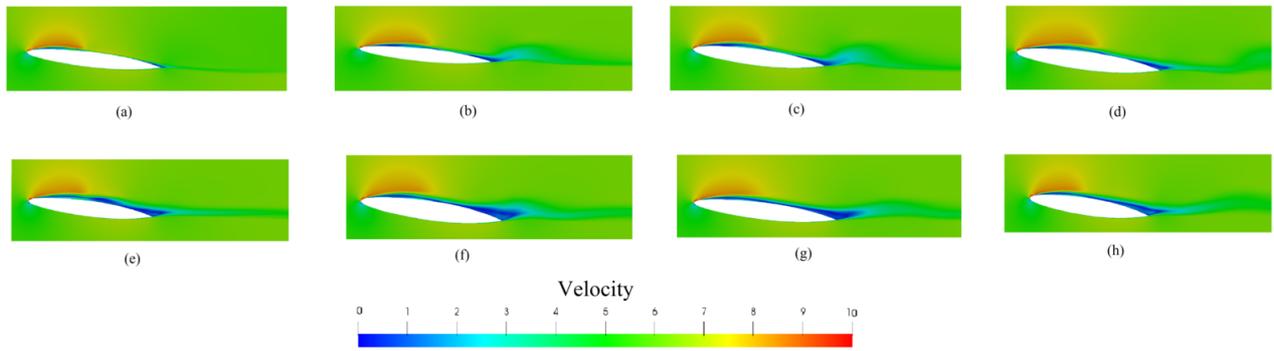

**Figure 3.** Instantaneous velocity structures around the hydrofoil with CG. The time step between images is 19 ms.

## 4. Results and discussions

Unsteady cloud cavitating flow around the CAV2003 benchmark hydrofoil with and without CGs are presented here. The geometry of this hydrofoil was offered in CAV2003 workshop in Osaka [22]. The alternate generation and shedding of the large cloud cavity in the period causes to the velocity fluctuations on suction side and main oscillation with high amplitude. There is also a secondary oscillation with smaller amplitude and higher frequency in the unsteady cloud cavitation which induces by the alternate generation and shedding of the small vortexes or collapse process of small bubbles. Fig. 2 shows the instantaneous velocity contours in one oscillation cycle around the hydrofoil without CGs. According to Fig. 2, it is clear that the growth and collapse of the cavitating bubble changes the local flow structure which causes to turbulent velocity fluctuations in the unsteady cavitation regime on the suction side of the hydrofoil. Fig. 3 shows the instantaneous velocity structures in one oscillation cycle with the presence of CG. It can be seen from Fig. 3 (a-h) that the local flow structures have not any significant changes which may be due to stabilisation of the local boundary layer structure. The images show a stable turbulent velocity fluctuations in the unsteady cavitation regime on the suction side and at the trailing edge. The cyclic physics of cavitation becomes different using CG such that the cyclic process disappears and the cavitation phenomena becomes into a stable form with low-amplitude fluctuations. In the figures 4-8 the evolution of the velocity vectors and vortex structures are analyzed to show some mechanisms behind the shedding of the cloud cavity and cavitation unsteadiness on the suction side of the hydrofoil with and without CGs. Fig. 4a shows a vortex called 'Vortex 1' shedding near the trailing edge of the hydrofoil at the beginning of the oscillation cycle. The adverse pressure gradient at the cavity closure and the re-entrant jet occur after the sheet cavity reaches to its maximum length. This process results from another vortex called 'Vortex 2', see Fig. 4b. Eventually after the shedding of the cavity the 'Vortex 2' moves towards the trailing edge and becomes a larger vortex called 'Vortex 3', see Fig. 4c.

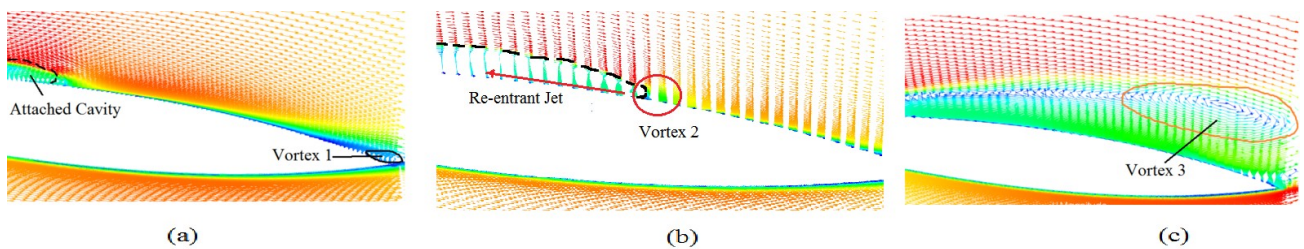

**Figure 4.** Different vortexes and inverse flow developed on suction side of the hydrofoil without CGs at different instants, Black line represents cavity shape. The time step between images is 19 ms.

Fig. 5 shows the evolution of the vortexes on suction side of hydrofoil behind CG at different instants. It can be seen that the local flow structures and the size of vortexes behind CG do not change significantly. According to instantaneous images of Fig. 6, it is clear that the local boundary layer and the vortex shedding to the trailing edge on suction side of hydrofoil were changed slightly and there are not any significant changes at the positions of the vortex core at the trailing edge of the hydrofoil with CG. Fig. 7 and 8 show the evolution of the velocity vectors at the leading edge on suction side without and with CGs in 3D view. It can be seen from Fig. 7 that the size of vortexes and the velocity vectors change during the oscillation cycles. The vortex structures continuously strengthened by high adverse pressure gradient and causes to generation of a re-entrant jet directed towards the leading edge. The intermittent growth and shrinkage in the boundary layer thickness and turbulent velocity fluctuations that induce boundary layer instabilities can be seen from Fig. 7. In contrast the velocity vectors and the shape of vortex structure behind the CG were changed insignificantly as shown in Fig. 8. Consequently, the high-amplitude of fluctuations of the lift and drag forces acting on the hydrofoil may be decreased. This may be obtained through the suppression or delay of the intermittent large cloud cavities that affect the fluctuations of the lift and drag forces of the hydrofoil. The time history of the velocity profiles at two positions in $x/c = 1.1$ and $x/c = 1.3$ from the trailing edge of the hydrofoil with and without CGs were shown in Fig. 9 and 10. It can be seen from the figures that with the modification of the boundary layer thickness on the suction side of the hydrofoil the velocity near to the hydrofoil wall and at the vicinity of the trailing edge were increased in comparison with the hydrofoil without CGs. The Fig. 10 shows that the instabilities of the velocity profile were stabilized using CG. This means that the velocity profile fluctuations in wake region were reduced in comparison with the hydrofoil without CGs. This reduction may be obtained with regard to the effects of the enhancement of the momentum exchanges with the flow external layer.

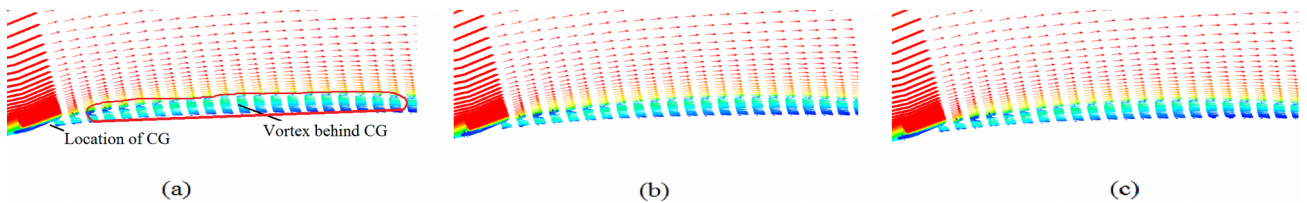

**Figure 5.** Vortex structure on suction side behind CG at different instants. The time step between images is 19 ms.

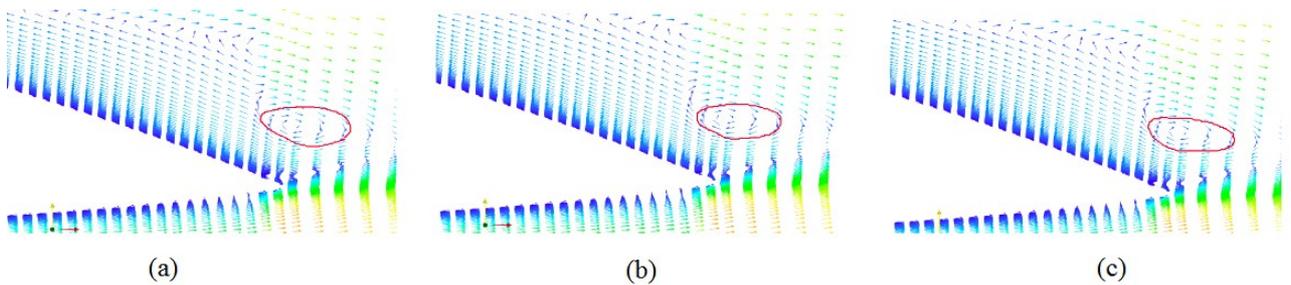

**Figure 6.** The vortex shedding to the trailing edge on suction side with CG at different instants. The time step between images is 19 ms.

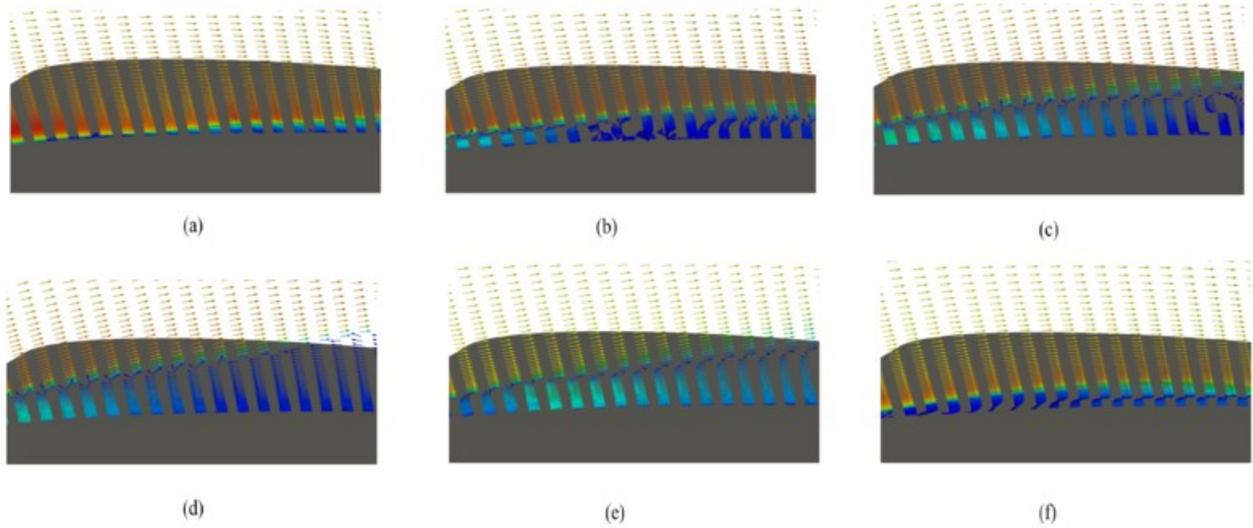

**Figure 7.** The evolution of the velocity vectors at the leading edge on suction side without CGs. The time step between images is 19 ms.

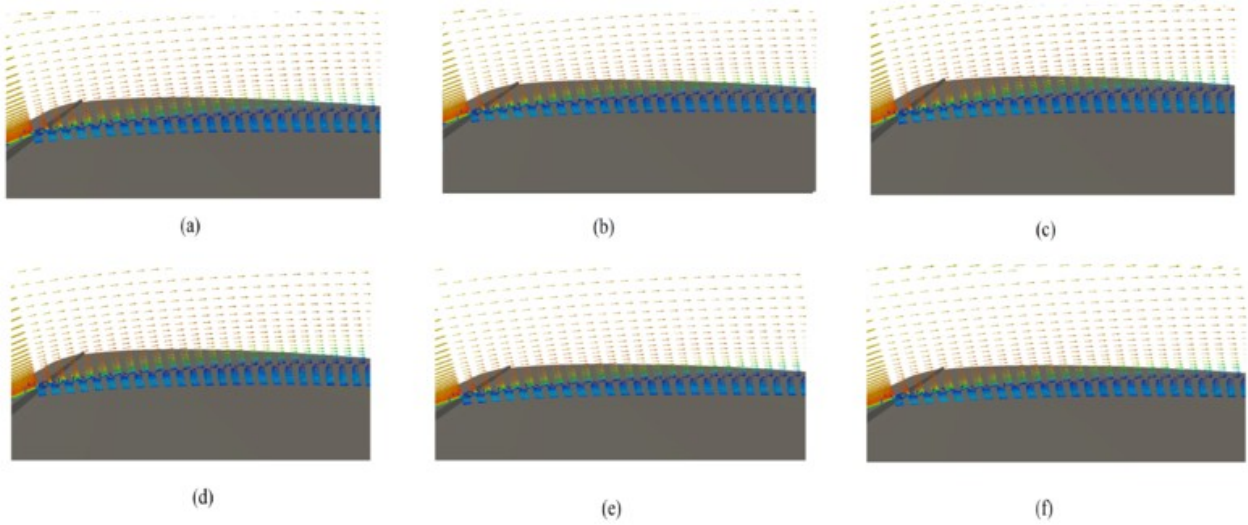

**Figure 8.** The evolution of the velocity vectors at the leading edge (behind the CG) on suction side with CG. The time step between images is 19 ms.

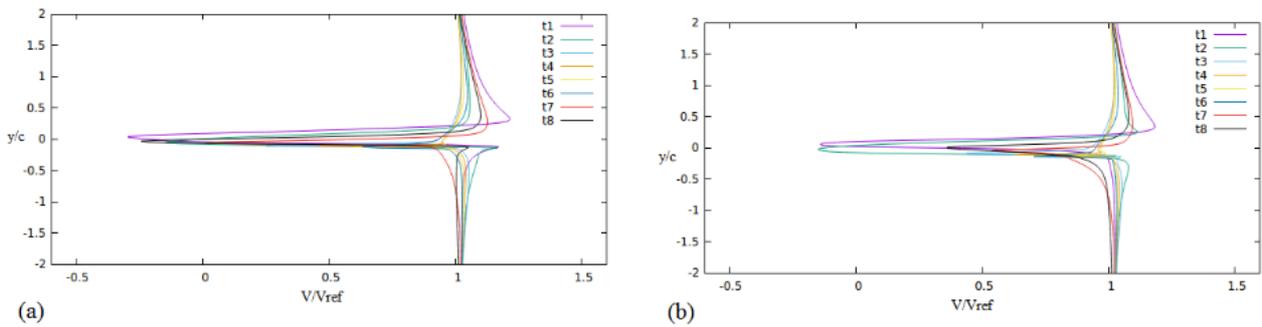

**Figure 9.** Time history of velocity profiles at two positions from the trailing edge of the hydrofoil without CGs. a) Position in x/c = 1.1, b) Position in x/c = 1.3. The time step between images is 19 ms.

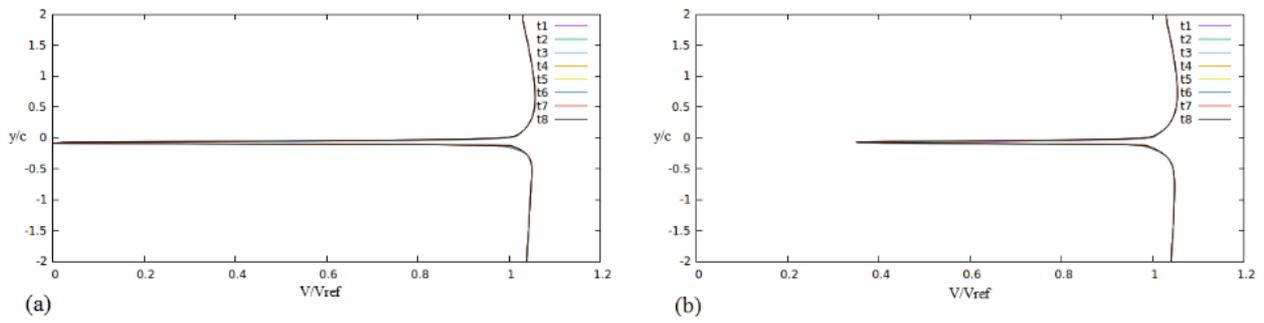

**Figure 10.** Time history of velocity profiles at two positions from the trailing edge of the hydrofoil with CG. a) Position in x/c = 1.1, b) Position in x/c = 1.3. The time step between images is 19 ms.

The evolution of the vorticity downstream from the trailing edge at x/c = 1.2, x/c = 1.4, x/c = 1.6, x/c = 1.8 and x/c = 2 are shown in Fig. 11. This figure shows that the vortex structures at the trailing edge of the hydrofoil were reduced that may cause a reduction in high frequency cavitation induced noise emitted due to the collapse of small cavity bubbles and vortexes at the trailing edge.

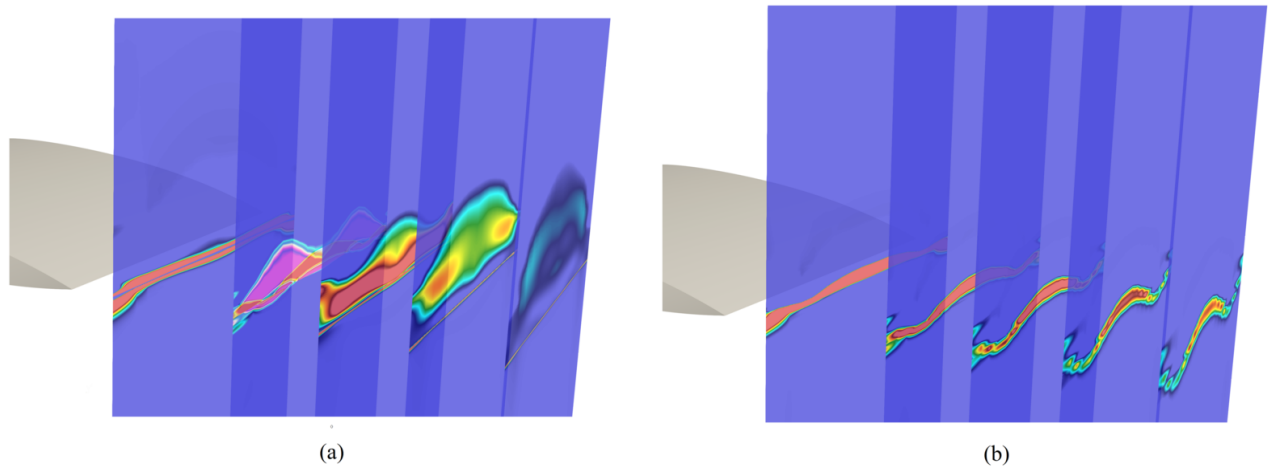

**Figure 11.** Iso-surface of vorticity in the wake region at different locations with ω = 500 [s −1 ], (a): Without CGs (b): With CG

## 5. Conclusions
Our results showed that the boundary layer instabilities may be stabilized and the turbulent velocity fluctuations can be reduced significantly using CG. With the stabilisation of the boundary layer and delays its separation on the hydrofoil surface the vortex structure of the cavity on the suction side and in the wake region were changed. Further on, the amplitude of the dominant frequency was decreased remarkably which may result from changes in the cavitation type from cloud to a quasi-steady cavitation. Using CG a reduction of the amplitude of primary and secondary oscillations can be reached. This means a significant reduction on vibration energy in low and high frequency bands may occurs.

## 6. Acknowledgement
The author is grateful to Dr. Javadi from flow control research lab for the fruitful discussion. The authors would like to thank the support of computer resources by MagnitUDE supercomputers of